\documentclass{article}
\usepackage{spconf}
\usepackage{amsmath}
\usepackage{amssymb}
\usepackage{graphicx}
\usepackage{hyperref}

\title{POLARIS: Training-Free Audio Fingerprinting with Saliency-Based Landmarks and Delaunay Grouping}
\name{Jiheng Li}
\address{Department of Computer Science, Vanderbilt University, Nashville, TN, 37235}
\begin{document}
\ninept
\maketitle
\begin{abstract}
This work presents POLARIS, a training-free audio fingerprinting system
that selects landmarks from a locally normalized saliency field and groups them into sparse fingerprints using Delaunay triangulation. To deal with query distortion, POLARIS adds fingerprints from two-hop Delaunay neighborhoods only at query time, without enlarging the reference index. An adaptive configuration applies this expansion only when the original fingerprints do not produce a confident match. We evaluate POLARIS on synthetic distortions from the public PEX Hard Medium benchmark, excluding queries with pitch or tempo shifts, and on a new benchmark of real re-recorded music. POLARIS achieves the best performance among the evaluated training-free methods on both benchmarks. On the real recordings, its adaptive configuration also outperforms the neural NMFP baseline with a comparable measured query time and a smaller logical reference payload. Code, dataset, and instructions for reproducing all experiments are available at \url{https://github.com/JihengLi/POLARIS.git}.
\end{abstract}
\section{Introduction}
\label{sec:intro}

Audio fingerprinting identifies an unknown audio clip, or query, by matching it against known reference recordings stored in a database  \cite{cano_review_2005,wang_industrial_2003}. Its applications include music recognition, broadcast monitoring, and copyright tracking  \cite{cano_review_2005,wang_industrial_2003}. In practice, queries may be compressed, mixed with noise, filtered, reverberated, or recorded through an acoustic channel. This study focuses on such distortions when global pitch and tempo are preserved. A practical system in this setting must remain reliable while keeping the reference index compact and query processing efficient.

Recent audio fingerprinting systems broadly follow learned and training free approaches. Neural methods such as NAFP, GraFPrint, and NMFP learn distortion-robust embeddings from augmented audio \cite{chang_neural_2021,bhattacharjee_grafprint_2025,araz_enhancing_2025}. Training-free systems instead commonly follow a landmark-and-hash workflow: they select sparse landmarks from a time-frequency representation, group nearby landmarks into fingerprints, and retrieve candidate recordings in the reference database \cite{wang_industrial_2003,noauthor_robust_2013,kim_robust_2014,six_olaf_2023}. 

The landmark detector varies across systems. Shazam, Audfprint, and OLAF use spectral peaks \cite{wang_industrial_2003,noauthor_robust_2013,six_olaf_2023}, whereas Kim et al. normalize and emphasize an MCLT spectrum before peak selection \cite{kim_robust_2014}. Landmark robustness therefore depends partly on how spectral prominence is defined.

Once landmarks have been detected, fingerprint construction determines which of them are encoded together. Representative systems commonly use local rules such as prescribed target regions and nearest-neighbor selection \cite{wang_industrial_2003,noauthor_robust_2013,six_olaf_2023,six_panako_2014,sonnleitner_quad-based_2014}. Delaunay triangulation provides another grouping rule: each triangular face defines a triplet fingerprint. It has been used in biometric fingerprint recognition \cite{bebis_fingerprint_1999,liang_robust_2007}, and disclosed in a patent as one possible grouping rule for spectrogram interest points \cite{sharifi_transformation_2014}. The patent describes ratio-based descriptors aggregated over time windows, but does not report a controlled empirical evaluation of Delaunay grouping for audio identification. 

Reference and query processing need not be symmetric. Seo retains
auxiliary information only for query matching
\cite{seo_asymmetric_2014}, while Schreiber and M{\"u}ller enlarge the
query fingerprint and subsample the reference index \cite{schreiber_accelerating_2014}. POLARIS uses a different form of asymmetry: the reference index stores fingerprints constructed from Delaunay faces, whereas the query also includes fingerprints from two-hop Delaunay neighborhoods.

Evaluation presents a further challenge. Synthetic distortions provide controlled and reproducible tests, but do not fully capture the effects of real playback equipment and acoustic environment. Recent work has therefore extended audio fingerprinting evaluation to broadcast and microphone-recorded audio \cite{araz_enhancing_2025,cortes_baf_2022}.

Against this background, the main contributions of POLARIS are:
\begin{itemize}
\item \textbf{Saliency-based landmark selection.} We propose a locally normalized saliency field and select its local maxima as landmarks in nonuniform time-frequency backgrounds.
\item \textbf{Delaunay landmark grouping with adaptive query-side expansion.} We construct fingerprints from Delaunay faces and add query fingerprints from two-hop Delaunay neighborhoods. The adaptive configuration skips this expansion when the initial match is already confident.
\item \textbf{Evaluation with real acoustic re-recordings.} We introduce a benchmark of real smartphone re-recordings of openly licensed music and complement it with pitch- and tempo-preserving distortions from the public PEX Hard Medium benchmark \cite{noauthor_pexesoaudio-fingerprinting-benchmark-toolkit_2026}.
\end{itemize}

\section{Method}
\label{sec:method}

\subsection{Saliency-based Landmark Selection}
\label{sec:landmark-selection}

Conventional landmark-based systems commonly detect landmarks as local maxima of spectrogram magnitude \cite{wang_industrial_2003,noauthor_robust_2013,six_olaf_2023}. Such local-maximum detection ensures that a selected landmark exceeds its neighbors, but it does not measure how large the difference is or account for variations in the local background. POLARIS therefore detects landmarks from a locally normalized saliency field rather than directly from spectrogram magnitude.

The saliency field combines a normalized contrast measure with a magnitude term. The contrast measure addresses the two limitations above: it first measures how far each bin lies above its local background and then scales this difference according to the contrast variation in the surrounding region. The magnitude term favors stronger spectral components when normalized contrast is similar.

\textbf{Normalized contrast.} Let $X\in\mathbb{R}^{F\times T}$ denote the log-magnitude spectrogram, normalized to have a maximum of 80. To measure the difference from the local background, we first smooth the spectrogram and subtract a broader background estimate:
\begin{equation}
X_s=\mathcal{G}_s(X), \qquad
C=X_s-\mathcal{G}_b(X_s).
\label{eq:local-contrast}
\end{equation}
Here, $\mathcal{G}_s$ and $\mathcal{G}_b$ are Gaussian smoothing operators at fine and broad scales, respectively. The value of $C$ still depends on how much the spectrum normally varies in that region. We therefore normalize it by the surrounding contrast energy:
\begin{equation}
N=
\frac{C}
{\sqrt{\mathcal{G}_n(C^2)+\epsilon^2}}.
\label{eq:normalized-contrast}
\end{equation}
Here, $\mathcal{G}_n$ is a Gaussian smoothing operator and $\epsilon$ prevents division by zero.

\textbf{Magnitude term.} Finally, we combine the normalized contrast with the magnitude term:
\begin{equation}
S=\beta N+\gamma[X_s-\tau]_+ .
\label{eq:saliency}
\end{equation}
Here, $[x]_+=\max(x,0)$, $\tau$ is the magnitude threshold, and
$\beta$ and $\gamma$ control the two contributions.

\textbf{Selection.} POLARIS detects local maxima directly on the saliency field $S$, discards candidates with $S\leq5$, and ranks the remaining candidates by $S$. A 2-s sliding-window rate controller retains approximately 22 landmarks per second. The retained landmarks are used for Delaunay fingerprint construction.

\subsection{Delaunay Fingerprint Construction}
\label{sec:fingerprint-construction}

POLARIS uses Delaunay triangulation to group landmarks into fingerprints.

Before triangulation, we normalize the landmark coordinates because
time frames and frequency bins use different units. Each landmark
$(f_i,t_i)$ is mapped to $\mathbf{z}_i=(f_i/r_f,t_i/r_t)$, where
$r_f$ and $r_t$ control the relative contribution of frequency and
time to the geometry. All triangle measurements below are computed
in this normalized plane. 

\begin{figure}[t]
\centering
\includegraphics[width=\columnwidth] {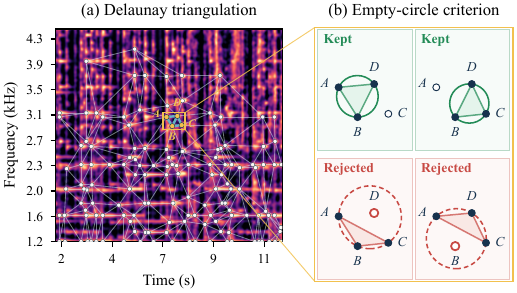}
\caption{Delaunay fingerprint construction. (a) Selected landmarks are organized by Delaunay triangulation on the saliency field. (b) The four possible triplets are evaluated using the empty-circle criterion.}
\label{fig:delaunay}
\end{figure}

For each recording, the triangulation is computed over all selected
landmarks. It partitions their convex hull into triangular faces with non-overlapping interiors (Fig.~\ref{fig:delaunay}a). Each face satisfies the empty-circle criterion: the interior of each triangle's circumcircle contains no other landmark (Fig.~\ref{fig:delaunay}b). Each triangular face forms a candidate fingerprint. The grouping
is therefore determined by the geometry of the landmark set rather
than by a prescribed target region around each anchor.

A useful property of Delaunay triangulation is that the number of triangles grows linearly with the number of landmarks. For $n$ landmarks in general position, with $n_h$ landmarks on the convex hull, the triangulation contains $2n-2-n_h<2n$ triangles \cite{de_berg_delaunay_2008}. POLARIS therefore considers fewer than $2n$ candidate reference fingerprints rather than all $\binom{n}{3}$ possible combinations of three landmarks.

Before hashing, POLARIS discards degenerate triangles and triangles spanning an overly large time-frequency region using bounds on their time span, normalized area, longest edge, and circumradius.

\textbf{Fingerprint hashing.} For each retained triangle, the vertices are ordered by time as $(f_1,t_1)$, $(f_2,t_2)$, and $(f_3,t_3)$. The fingerprint encodes the absolute frequency of the first vertex and the relative time-frequency offsets of the other two vertices. Following standard landmark hashing \cite{wang_industrial_2003,six_panako_2014}, these five quantities are quantized to form the hash
\begin{equation}
\mathbf{h}_{\Delta}
=
\mathcal{Q}\big(
\left[
f_1,\,
f_2-f_1,\,
t_2-t_1,\,
f_3-f_1,\,
t_3-t_1
\right]
\big),
\label{eq:triangle-hash}
\end{equation}
where $\mathcal{Q}$ denotes component-wise quantization. $t_1$ is stored alongside the hash for temporal alignment during matching.

\subsection{Query-side Expansion and Matching}
\label{sec:query-expansion}

In a query recording, noise and other distortions may shift landmark positions or cause additional landmarks to be detected. These changes
can alter the Delaunay triangulation, so even when all three landmarks of a reference fingerprint are still detected, they may no longer form a face together. 

\textbf{Two-hop neighborhood expansion.} To address this problem, POLARIS applies a two-hop neighborhood expansion only on the query side, while leaving the reference index unchanged. For each query landmark, it considers landmarks that occur later in time and can be reached by following at most two edges in the query Delaunay graph. It combines the current landmark with two of these neighbors to form an additional query fingerprint, without requiring the three landmarks to form a Delaunay face. The expansion is limited to the 12 nearest of these neighbors and 24 additional fingerprints per landmark.

\textbf{Matching.}
POLARIS uses conventional hash matching and time-offset voting \cite{wang_industrial_2003}. Each query hash retrieves reference entries with the same value. To reduce misses near quantization boundaries, POLARIS also looks up hashes obtained by increasing or decreasing one of the five quantized values by one. Each match votes for its track at the time difference between the matched reference and query fingerprints, so correct matches should cluster at a consistent offset. Matches from hashes found in fewer reference tracks receive greater weight, reducing the influence of common hashes. The strongest weighted cluster gives the track score and estimated query offset.

\textbf{Adaptive query expansion.}
Many queries can be matched confidently without two-hop expansion. To reduce retrieval cost, POLARIS uses a two-stage procedure. It first matches fingerprints constructed from the query Delaunay faces, without expansion. If the top track has at least 8 matching hashes and a score margin of at least 0.40 over the second-ranked track, the result is returned immediately. Otherwise, POLARIS adds the two-hop fingerprints and performs matching again. The margin is $(s_1 -s_2)/s_1$, where $s_1$ and $s_2$ are the two highest track scores.

\section{Experimental Setup}
\label{sec:experiment}

\subsection{Datasets}
\label{sec:datasets}

We evaluate on one public benchmark with synthetic distortions and one benchmark with real acoustic re-recordings.

\textbf{PEX.} We use the PEX Hard Medium dataset \cite{noauthor_pexesoaudio-fingerprinting-benchmark-toolkit_2026}, which contains 953 reference tracks. Because POLARIS does not model global pitch or tempo changes, we retain only annotations whose tempo field is empty or 100 and whose pitch field is empty or 0. We extract and evaluate each retained annotated segment as a separate query, yielding 791 queries covering compression, filtering, noise, gain variation, echo, and combinations of these distortions. 

\textbf{SD-RR.}
We construct the Song Describer Real Re-recording benchmark (SD-RR) \cite{li_sd-rr_2026} from openly licensed music in the Song Describer collection \cite{manco_song_2023}. After excluding tracks whose licenses prohibit derivative use and one track without enough usable audio for three non-overlapping excerpts, 496 of the original 706 tracks remain and form the reference collection. Using a fixed random seed (20260819), we select three non-overlapping 10-s excerpts from each track, yielding 1,488 queries.

Before playback, each excerpt is normalized to $-20$\,dBFS RMS, with peaks limited to $-1$\,dBFS. To determine the ground-truth offset of each query, the three excerpts from each track are placed at known positions in a single playback sequence together with short synchronization chirps. We re-record these sequences continuously in batches and detect the chirps to align each phone recording with its corresponding playback sequence. Combining this alignment with the original location of each excerpt gives its ground-truth reference offset. A separate spectral alignment check agreed with the chirp-based offsets to within 0.1 s for 99.1\% of the queries.

\textbf{Recording setup.}
The built-in loudspeakers of a 14-inch MacBook Pro (2023) play each sequence at a fixed system-volume setting of 40\%. An iPhone 14 Pro Max placed approximately 1.6\,m away records the playback using Apple Voice Memos. We record 460 tracks in a residential apartment and the remaining 36 outdoors.

\subsection{Compared Systems}
\label{sec:compared-systems}

\textbf{POLARIS configurations.} We report three query-processing modes of POLARIS. POLARIS-O uses fingerprints constructed from the original Delaunay faces. POLARIS-F always combines these with the two-hop query fingerprints. POLARIS-A begins with POLARIS-O and adds the two-hop fingerprints only when the leading match does not satisfy the confidence criterion. 

\textbf{Baselines.}
We compare POLARIS with Audfprint \cite{noauthor_robust_2013}, Panako v2.1 \cite{six_panako_2014,six_panako_2021}, OLAF v2.0.10 \cite{six_olaf_2023}, and NMFP \cite{araz_enhancing_2025}. Audfp-M and
OLAF are configured to approximately match the logical reference
payload of POLARIS; their settings are selected from reference payload
alone, without using query results. Audfp-Q uses the Audfp-M reference index but generates more query fingerprints, approximately matching the query payload of POLARIS-F on SD-RR. This tests whether increasing the Audfprint query budget can match POLARIS accuracy. For closed-set evaluation, Audfp-M and OLAF return a candidate whenever at least one match is available, while Panako's rejection thresholds are set to their minimum values. NMFP uses the authors' pretrained NMFP-Triplet checkpoint without retraining; retrieval uses exhaustive embedding search followed by temporal alignment. Full configurations, software revisions, and exact commands are provided with the released code. 

\textbf{Ablations.}
The comparison between POLARIS-O and POLARIS-F evaluates two-hop query
expansion. We additionally evaluate two controlled variants on SD-RR. The first is a POLARIS-F variant that replaces saliency maxima with spectrogram-magnitude maxima while keeping the remaining pipeline unchanged. The second is a POLARIS-O variant that uses the same saliency landmarks and matcher but replaces Delaunay faces with target-region triplets adapted from OLAF \cite{six_olaf_2023}. Its reference and query payloads are approximately matched to POLARIS-O.

\begin{table}[t]
\centering
\caption{Main POLARIS parameters.}
\label{tab:polaris-parameters}

\setlength{\tabcolsep}{3pt}
\begin{tabular}{@{}p{0.29\columnwidth}p{0.65\columnwidth}@{}}
\hline
\textbf{Component} & \textbf{Setting} \\
\hline
Audio
& 40~kHz; Hann STFT, 2560/1280-sample window/hop; 0.78-5.47~kHz analysis band; four query STFT frame offsets spaced by 320 samples \\

Saliency
& $\beta=3.0$, $\gamma=0.015$, $\tau=10$, $\epsilon=0.75$; $\sigma_s=(1,1.5)$, $\sigma_b=(8,16)$, $\sigma_n=(6,12)$ \\

Landmark selection
& $5\times7$ local-maximum neighborhood on $S$ \\

Delaunay grouping
& $r_f=10$, $r_t=14$; time span 1-63 frames; normalized area $\geq0.025$; edge length $\leq6$; circumradius $\leq4$ \\

Hash encoding
& Step sizes: 2 frequency bins for absolute frequency;
3 for frequency differences; 2 frames for time differences \\
\hline
\end{tabular}
\end{table}

\subsection{Metrics}
\label{sec:metrics}

\textbf{Retrieval.} We report track Top-1 accuracy. A query is counted as correct when the top-ranked result is the ground-truth track. No-candidate outputs produced by Panako are counted as incorrect, not excluded from evaluation. For SD-RR, we also report the track-and-offset accuracy, which additionally requires the estimated time offset to be within $0.1$~s of the annotated offset. 

\textbf{Payload and query time.} Payload measures logical fingerprint data rather than serialized database storage. For POLARIS, Audfprint, and OLAF, a reference hash posting occupies 16 bytes and a query hash record occupies 12 bytes. A Panako reference posting occupies 20 bytes, while NMFP reference and query embedding records occupy 520 and 516 bytes, respectively. Database-specific overhead, shared track metadata, and model weights are excluded. Query payload is reported only when the evaluated interface exposes the corresponding logical record count. The Panako interface does not expose its query-fingerprint count, so its query payload cannot be computed. Query time covers processing from query loading to the returned ranking, with all systems measured on the same hardware. The reported values describe the end-to-end performance of the evaluated implementations.

\subsection{Implementation Details}
\label{sec:implementation}

All systems were run on an Apple M2 Max CPU with 32 GB RAM. Table~\ref{tab:polaris-parameters} lists the main parameters. We used the PEX Hard Small test split for development and froze all settings before evaluating on PEX Hard Medium and SD-RR.

PEX Hard Small and Hard Medium share 12 reference-track identifiers, but only nine of the queries in PEX Hard Medium use these tracks. Excluding these queries changes the track Top-1 accuracy of all three POLARIS configurations by no more than 0.08 percentage points.

\section{Results and Discussion}
\label{sec:results}

\subsection{Retrieval Accuracy}

Table~\ref{tab:main-results} summarizes the main results. On the evaluated PEX subset, all three POLARIS configurations outperform all evaluated training-free methods. Two-hop expansion improves over POLARIS-O, while POLARIS-A remains within 0.13 percentage points of POLARIS-F. NMFP retains the highest PEX accuracy.

On SD-RR, POLARIS-O, which uses only fingerprints from the original Delaunay faces, achieves higher track accuracy than every evaluated baseline. POLARIS-A and POLARIS-F obtain the same highest track accuracy, while POLARIS-F gives higher track-and-offset accuracy.

Increasing the Audfprint query budget improves accuracy on SD-RR but reduces it on PEX. In contrast, two-hop expansion improves POLARIS-O on both benchmarks. At approximately the same SD-RR query payload, POLARIS-F is also more accurate and faster than Audfp-Q in the evaluated implementations. Thus, increasing the Audfprint query budget does not reproduce the gains of POLARIS.

\begin{table}[t]
\centering
\caption{Top-1 accuracy on the evaluated PEX subset and SD-RR. Track+off. additionally requires an estimated offset within 0.1~s of the ground truth. Bold and underlined entries are the best and second-best accuracies per column. Logical payload and mean query time are reported for SD-RR. Panako produces no matching evidence for 832 SD-RR queries; these outputs are counted as incorrect.}
\label{tab:main-results}
\setlength{\tabcolsep}{1.5pt}
\begin{tabular*}{\columnwidth}{
@{\extracolsep{\fill}}lcccccc@{}
}
\hline
& \textbf{PEX}
& \multicolumn{2}{c}{\textbf{SD-RR accuracy}}
& \multicolumn{3}{c}{\textbf{SD-RR cost}} \\
\cline{2-2}\cline{3-4}\cline{5-7}
\textbf{Method}
& \textbf{Track}
& \textbf{Track}
& \textbf{Track+off.}
& \textbf{Ref.}
& \textbf{Query}
& \textbf{Time} \\
& \textbf{(\%)}
& \textbf{(\%)}
& \textbf{(\%)}
& \textbf{MiB/h}
& \textbf{MiB/q}
& \textbf{s/q} \\
\hline
Audfp-M
& 90.64 & 61.02 & 50.87 & 1.88 & 0.011 & 0.11 \\

Audfp-Q
& 88.12 & 67.20 & 60.35 & 1.88 & 1.434 & 1.44 \\

Panako
& 73.70 & 4.84 & 4.37 & 1.74 & -- & 0.92 \\

OLAF
& 89.89 & 24.40 & 18.95 & 2.09 & 0.002 & 0.09 \\

NMFP
& \textbf{98.99} & 74.33 & 71.17 & 3.54 & 0.009 & 0.33 \\

\hline
POLARIS-O
& 93.30
& \underline{79.57}
& 70.16
& 1.89
& 0.152
& 0.15 \\

POLARIS-A
& 93.93
& \textbf{84.61}
& \underline{75.34}
& 1.89
& 0.538
& 0.30 \\

POLARIS-F
& \underline{94.06}
& \textbf{84.61}
& \textbf{76.21}
& 1.89
& 1.473
& 0.68 \\
\hline
\end{tabular*}
\end{table}

\subsection{Reference Payload and Query Cost}
Despite the accuracy gains above, all three POLARIS configurations use
the same sparse reference index. Its payload is comparable to that of the evaluated training-free baselines and smaller than that of NMFP. Instead of storing more fingerprints for every reference track, POLARIS generates additional fingerprints only while processing a query. These
fingerprints are temporary and do not add to the reference storage that
grows with the size of the collection.

The additional query processing increases retrieval time from
POLARIS-O to POLARIS-F, but the measured times remain competitive.
POLARIS-O is faster than NMFP while achieving higher SD-RR track
accuracy, and POLARIS-A exceeds NMFP on both accuracy measures with
essentially the same mean query time. On SD-RR, adaptive processing reduces query payload by 63.5\% and mean query time by 55.9\% relative to POLARIS-F while preserving track accuracy. The relative time saving is larger on PEX: more than 90\% of the queries stop after the first stage, reducing mean query time by 72.3\% relative to POLARIS-F with only a 0.13-percentage-point loss in track accuracy.

\subsection{Ablation Study}

Table~\ref{tab:ablation} evaluates the three main design choices in
POLARIS. Replacing saliency maxima with spectrogram-magnitude maxima produces the largest accuracy reduction. This result shows that saliency-based landmark selection contributes substantially to robustness.

The target-region control evaluates the fingerprint grouping rule.
With approximately matched reference and query payloads, it is less
accurate than POLARIS-O on both measures. This controlled comparison
supports Delaunay grouping over the evaluated target-region alternative
without attributing the difference to a larger fingerprint budget.

Finally, POLARIS-F improves both accuracy measures over POLARIS-O,
showing that two-hop expansion recovers useful query matches beyond the
original Delaunay faces. Taken together, the controls support all three
parts of the method: saliency-based landmark selection, Delaunay
fingerprint construction, and query-side neighborhood expansion.

\begin{table}[t]
\centering
\caption{Ablation results on SD-RR. Accuracy is reported in percent; best accuracies are shown in bold. Query payload and processing time are averaged over the 10-s queries.}
\label{tab:ablation}
\setlength{\tabcolsep}{3pt}
\begin{tabular}{@{}lcccc@{}}
\hline
& \multicolumn{2}{c}{\textbf{Top-1 accuracy}}
& \textbf{Query}
& \textbf{Time} \\
\cline{2-3}\cline{4-4}\cline{5-5}
\textbf{Variant}
& \textbf{Track}
& \textbf{Track+off.}
& \textbf{MiB/q}
& \textbf{s/q} \\
\hline
Magnitude maxima
& 62.03
& 57.46
& 1.845
& 2.51 \\

Target-region triplets
& 71.71
& 65.19
& 0.138
& 0.41 \\

POLARIS-O
& 79.57 & 70.16 & 0.152 & 0.15 \\

POLARIS-F
& \textbf{84.61}
& \textbf{76.21}
& 1.473
& 0.68 \\
\hline
\end{tabular}
\end{table}

\section{Conclusions and Future Work}
\label{sec:conclusions}

POLARIS combines saliency-based landmark selection, Delaunay grouping, and query-side expansion in a training-free audio fingerprinting system. On SD-RR, the controlled variants show improvements from saliency-based selection, Delaunay grouping, and two-hop expansion over their evaluated alternatives. All three POLARIS configurations outperform the evaluated training-free baselines on both benchmarks while sharing the same sparse reference index. On SD-RR, POLARIS-A also exceeds NMFP in both accuracy measures with a comparable mean query time and a smaller reference representation. These results show that query-side geometric expansion can improve robustness without enlarging the stored reference index.

This study has several limitations. First, the evaluation is closed-set and focuses on distortions that preserve global pitch and tempo. Future work will extend the geometry to cover pitch and tempo changes and evaluate open-set identification against larger reference collections. Second, SD-RR was recorded with one laptop and one smartphone in a limited set of environments. Future versions can include additional rooms, devices, and outdoor conditions. Third, the reported query processing times depend on programming language, database backend, and optimization level and should not be interpreted as intrinsic algorithmic speed. The current POLARIS implementation has not been extensively optimized for latency; future work can improve query fingerprint generation and index lookup. Finally, NMFP was trained on FMA-derived audio, while PEX is also constructed from FMA and therefore does not provide a cross-corpus test for this baseline. SD-RR provides such a complement, but broader evaluations are needed to compare the generalization of learned and training-free methods.

\vfill\pagebreak

\section{Compliance with Ethical Standards}
\label{sec:ethical-standards}
This study involved no human participants or animal subjects and did not require ethical approval.

\section{Acknowledgements}
\label{sec:acknowledgements}
No funding was received for conducting this study. The author has no relevant financial or nonfinancial interests to disclose.

\bibliographystyle{IEEEbib}
\bibliography{refs}

@inproceedings{cortes_baf_2022,
    address = {Bengaluru, India},
    title = {{BAF}: {An} audio fingerprinting dataset for broadcast monitoring},
    shorttitle = {{BAF}},
    url = {https://zenodo.org/records/7343030},
    doi = {10.5281/zenodo.7343030},
    urldate = {2026-08-24},
    booktitle = {Proceedings of the 23rd International Society for Music Information Retrieval Conference},
    publisher = {ISMIR},
    author = {Cortès, Guillem and Ciurana, Alex and Molina, Emilio and Miron, Marius and Meyers, Owen and Six, Joren and Serra, Xavier},
    month = dec,
    year = {2022},
    pages = {908--916},
}

@inproceedings{araz_enhancing_2025,
    address = {Daejeon, South Korea},
    title = {Enhancing {Neural} {Audio} {Fingerprint} {Robustness} to {Audio} {Degradation} for {Music} {Identification}},
    url = {https://zenodo.org/records/17811394},
    doi = {10.5281/zenodo.17811394},
    urldate = {2026-08-24},
    booktitle = {Proceedings of the 26th International Society for Music Information Retrieval Conference},
    publisher = {ISMIR},
    author = {Araz, Recep Oguz and Cortès-Sebastià, Guillem and Molina, Emilio and Serrà, Joan and Serra, Xavier and Mitsufuji, Yuki and Bogdanov, Dmitry},
    month = sep,
    year = {2025},
    pages = {399--406},
}

@article{cano_review_2005,
    title = {A {Review} of {Audio} {Fingerprinting}},
    volume = {41},
    issn = {0922-5773},
    url = {https://doi.org/10.1007/s11265-005-4151-3},
    doi = {10.1007/s11265-005-4151-3},
    language = {en},
    number = {3},
    urldate = {2026-08-24},
    journal = {Journal of VLSI signal processing systems for signal, image and video technology},
    author = {Cano, Pedro and Batlle, Eloi and Kalker, Ton and Haitsma, Jaap},
    month = nov,
    year = {2005},
    pages = {271--284},
}

@inproceedings{wang_industrial_2003,
    address = {Baltimore, United States},
    title = {An {Industrial-Strength} {Audio} {Search} {Algorithm}},
    url = {https://zenodo.org/records/1416340},
    doi = {10.5281/zenodo.1416340},
    urldate = {2026-08-24},
    booktitle = {Proceedings of the 4th International Conference on Music Information Retrieval},
    publisher = {ISMIR},
    author = {Wang, Avery},
    month = oct,
    year = {2003},
}

@article{six_olaf_2023,
    title = {Olaf: a lightweight, portable audio search system},
    volume = {8},
    issn = {2475-9066},
    shorttitle = {Olaf},
    url = {https://joss.theoj.org/papers/10.21105/joss.05459},
    doi = {10.21105/joss.05459},
    language = {en},
    number = {87},
    urldate = {2026-08-24},
    journal = {Journal of Open Source Software},
    author = {Six, Joren},
    month = jul,
    year = {2023},
    pages = {5459},
}

@misc{noauthor_robust_2013,
    author = {Ellis, Dan},
    title = {Robust {Landmark}-{Based} {Audio} {Fingerprinting}},
    url = {https://www.ee.columbia.edu/~dpwe/LabROSA/matlab/fingerprint/},
    howpublished = {\url{https://www.ee.columbia.edu/~dpwe/LabROSA/matlab/fingerprint/}},
    urldate = {2026-08-24},
    year = {2009},
}

@inproceedings{six_panako_2014,
    address = {Taipei, Taiwan},
    title = {Panako - {A} {Scalable} {Acoustic} {Fingerprinting} {System} {Handling} {Time}-{Scale} and {Pitch} {Modification}},
    url = {https://zenodo.org/records/1416190},
    doi = {10.5281/zenodo.1416190},
    urldate = {2026-08-24},
    booktitle = {Proceedings of the 15th International Society for Music Information Retrieval Conference},
    publisher = {ISMIR},
    author = {Six, Joren and Leman, Marc},
    month = oct,
    year = {2014},
    pages = {259--264},
}

@inproceedings{sonnleitner_quad-based_2014,
    title = {Quad-{Based} {Audio} {Fingerprinting} {Robust} to {Time} and {Frequency} {Scaling}},
    issn = {2413-6689},
    url = {https://dafx.de/paper-archive/details/6VJe_eYTwVo2jfIHzOkTDQ},
    urldate = {2026-08-24},
    author = {Sonnleitner, Reinhard and Widmer, Gerhard},
    year = {2014},
    booktitle = {Proceedings of the 17th International Conference on Digital Audio Effects ({DAFx-14})},
    pages = {173--180},
}

@article{seo_asymmetric_2014,
    title = {An {Asymmetric} {Matching} {Method} for a {Robust} {Binary} {Audio} {Fingerprinting}},
    volume = {21},
    issn = {1558-2361},
    url = {https://ieeexplore.ieee.org/document/6758355},
    doi = {10.1109/LSP.2014.2310237},
    number = {7},
    urldate = {2026-08-24},
    journal = {IEEE Signal Processing Letters},
    author = {Seo, Jin S.},
    month = jul,
    year = {2014},
    pages = {844--847},
}

@article{schreiber_accelerating_2014,
    title = {Accelerating {Index}-{Based} {Audio} {Identification}},
    volume = {16},
    issn = {1941-0077},
    url = {https://ieeexplore.ieee.org/document/6802350},
    doi = {10.1109/TMM.2014.2318517},
    number = {6},
    urldate = {2026-08-24},
    journal = {IEEE Transactions on Multimedia},
    author = {Schreiber, Hendrik and Müller, Meinard},
    month = oct,
    year = {2014},
    pages = {1654--1664},
}

@inproceedings{chang_neural_2021,
    title = {Neural {Audio} {Fingerprint} for {High}-{Specific} {Audio} {Retrieval} {Based} on {Contrastive} {Learning}},
    issn = {2379-190X},
    url = {https://ieeexplore.ieee.org/document/9414337},
    doi = {10.1109/ICASSP39728.2021.9414337},
    urldate = {2026-08-21},
    booktitle = {{ICASSP} 2021 - 2021 {IEEE} {International} {Conference} on {Acoustics}, {Speech} and {Signal} {Processing} ({ICASSP})},
    author = {Chang, Sungkyun and Lee, Donmoon and Park, Jeongsoo and Lim, Hyungui and Lee, Kyogu and Ko, Karam and Han, Yoonchang},
    month = jun,
    year = {2021},
    pages = {3025--3029},
}

@misc{noauthor_pexesoaudio-fingerprinting-benchmark-toolkit_2026,
    title = {Audio Fingerprinting Benchmark Toolkit},
    copyright = {MIT},
    url = {https://github.com/Pexeso/audio-fingerprinting-benchmark-toolkit},
    urldate = {2026-08-24},
    author = {{Pexeso}},
    howpublished = {GitHub repository, \url{https://github.com/Pexeso/audio-fingerprinting-benchmark-toolkit}},
    note = {Accessed: Aug. 24, 2026},
}

@misc{manco_song_2023,
    title = {The {Song} {Describer} {Dataset}: a {Corpus} of {Audio} {Captions} for {Music}-and-{Language} {Evaluation}},
    shorttitle = {The {Song} {Describer} {Dataset}},
    url = {http://arxiv.org/abs/2311.10057},
    doi = {10.48550/arXiv.2311.10057},
    urldate = {2026-08-27},
    publisher = {arXiv},
    author = {Manco, Ilaria and Weck, Benno and Doh, SeungHeon and Won, Minz and Zhang, Yixiao and Bogdanov, Dmitry and Wu, Yusong and Chen, Ke and Tovstogan, Philip and Benetos, Emmanouil and Quinton, Elio and Fazekas, György and Nam, Juhan},
    month = nov,
    year = {2023},
    note = {arXiv:2311.10057 [cs.SD]},
}

@inproceedings{six_panako_2021,
    title = {Panako 2.0: updates for an acoustic fingerprinting system},
    copyright = {info:eu-repo/semantics/openAccess},
    shorttitle = {Panako 2.0},
    url = {http://hdl.handle.net/1854/LU-8726851},
    language = {eng},
    urldate = {2026-08-28},
    booktitle = {{ISMIR} 2021 {Late}-breaking demo contributions},
    author = {Six, Joren},
    year = {2021},
}

@inproceedings{bhattacharjee_grafprint_2025,
    title = {{GraFPrint}: {A} {GNN}-{Based} {Approach} for {Audio} {Identification}},
    issn = {2379-190X},
    shorttitle = {{GraFPrint}},
    url = {https://ieeexplore.ieee.org/document/10888557},
    doi = {10.1109/ICASSP49660.2025.10888557},
    urldate = {2026-08-23},
    booktitle = {{ICASSP} 2025 - 2025 {IEEE} {International} {Conference} on {Acoustics}, {Speech} and {Signal} {Processing} ({ICASSP})},
    author = {Bhattacharjee, Aditya and Singh, Shubhr and Benetos, Emmanouil},
    month = apr,
    year = {2025},
    pages = {1--5},
}

@inproceedings{bebis_fingerprint_1999,
    title = {Fingerprint identification using {Delaunay} triangulation},
    url = {https://ieeexplore.ieee.org/document/810315},
    doi = {10.1109/ICIIS.1999.810315},
    urldate = {2026-08-29},
    booktitle = {Proceedings 1999 {International} {Conference} on {Information} {Intelligence} and {Systems} ({Cat}. {No}.{PR00446})},
    author = {Bebis, G. and Deaconu, T. and Georgiopoulos, M.},
    month = oct,
    year = {1999},
    pages = {452--459},
}

@misc{sharifi_transformation_2014,
    author = {Sharifi, Matthew and Ioffe, Sergey and Yagnik, Jay and Postelnicu, Gheorghe and Roblek, Dominik and Tzanetakis, George},
    title = {Transformation invariant media matching},
    howpublished = {U.S. Patent 8738633 B1},
    month = may,
    year = {2014},
    url = {https://patents.google.com/patent/US8738633B1/en}
}

@article{kim_robust_2014,
    title = {Robust {Audio} {Fingerprinting} {Method} {Using} {Prominent} {Peak} {Pair} {Based} on {Modulated} {Complex} {Lapped} {Transform}},
    volume = {36},
    copyright = {© 2014 ETRI},
    issn = {2233-7326},
    url = {https://onlinelibrary.wiley.com/doi/abs/10.4218/etrij.14.0113.1405},
    doi = {10.4218/etrij.14.0113.1405},
    language = {en},
    number = {6},
    urldate = {2026-08-29},
    journal = {ETRI Journal},
    author = {Kim, Hyoung-Gook and Kim, Jin Young},
    year = {2014},
    pages = {999--1007},
}

@article{liang_robust_2007,
    title = {A {Robust} {Fingerprint} {Indexing} {Scheme} {Using} {Minutia} {Neighborhood} {Structure} and {Low}-{Order} {Delaunay} {Triangles}},
    volume = {2},
    issn = {1556-6021},
    url = {https://ieeexplore.ieee.org/document/4380301},
    doi = {10.1109/TIFS.2007.910242},
    number = {4},
    urldate = {2026-08-30},
    journal = {IEEE Transactions on Information Forensics and Security},
    author = {Liang, Xuefeng and Bishnu, Arijit and Asano, Tetsuo},
    month = dec,
    year = {2007},
    pages = {721--733},
}

@incollection{de_berg_delaunay_2008,
    address = {Berlin, Heidelberg},
    title = {Delaunay {Triangulations}},
    isbn = {978-3-540-77974-2},
    url = {https://doi.org/10.1007/978-3-540-77974-2_9},
    doi = {10.1007/978-3-540-77974-2_9},
    language = {en},
    urldate = {2026-09-05},
    booktitle = {Computational {Geometry}: {Algorithms} and {Applications}},
    publisher = {Springer},
    author = {de Berg, Mark and Cheong, Otfried and van Kreveld, Marc and Overmars, Mark},
    year = {2008},
    pages = {191--218},
}

@misc{li_sd-rr_2026,
  author = {Li, Jiheng},
  title = {{SD-RR}: {Song Describer Real Re-recording Benchmark}},
  howpublished = {Zenodo, \url{https://doi.org/10.5281/zenodo.22169646}},
  month = aug,
  year = {2026},
  doi = {10.5281/zenodo.22169646},
  url = {https://zenodo.org/records/22169646}
}

\end{document}